\documentclass[aps,preprint,showpacs,floats,epsf,epsfig,nofootinbib,12pt]{revtex4}
\usepackage{graphicx}
\usepackage{epsfig}
\usepackage[dvips]{color}

\def\be{\begin{eqnarray}}
\def\en{\end{eqnarray}}

\begin{document}

\title{Study on radiative decays of $D^*_{sJ}(2860)$ and $D^*_{s1}(2710)$ into $D_s$ by means of LFQM}

\vspace{1cm}

\author{ Hong-Wei Ke$^{1}$   \footnote{khw020056@hotmail.com}, Jia-Hui Zhou$^{1}$ and
        Xue-Qian Li$^2$\footnote{lixq@nankai.edu.cn}  }

\affiliation{  $^{1}$ School of Science, Tianjin University, Tianjin 300072, China \\
  $^{2}$ School of Physics, Nankai University, Tianjin 300071, China }

\vspace{12cm}

\begin{abstract}

The observed resonance peak around 2.86 GeV has been carefully reexamined  by the LHCb collaboration and
it is found that under the peak there reside two states $D^*_{s1}(2860)$ and $D^*_{s3}(2860)$ which are considered as
$1^3D_1(c\bar s)$ and $1^3D_3(c\bar s)$ with slightly different masses and
total widths. Thus, the earlier assumption that the resonance $D^*_{s1}(2710)$
was a $1D$ state should not be right. We suggest to measure the partial widths of radiative decays of $D^*_{sJ}(2860)$ and $D^*_{s1}(2710)$ to confirm their quantum numbers.
We would consider $D^*_{s1}(2710)$ as $2^3S_1$ or  a pure $1^3D_1$ state, or their mixture and respectively calculate the corresponding branching ratios
as well as those of $D^*_{sJ}(2860)$. The future precise measurement would provide us information to help identifying the
structures of those resonances .


\end{abstract}

\maketitle

\section{Introduction}
Resonance
$D^*_s(2860)$
was experimentally observed \cite{Aubert:2009ah,Aubert:2006mh,Brodzicka:2007aa,Aaij:2012pc},
but its quantum number is still to be eventually identified because the ratio
$\Gamma(D^*_s(2860)\rightarrow D^*K)/\Gamma(D^*_s(2860)\rightarrow
DK)$ is not well understood \cite{Godfrey:2013aaa,Zhong:2009sk}. A careful reexamination
on the spectrum peak around 2.86 GeV recently has been
carried out by  the LHCb
collaboration  and it is found that
a spin-1 state and a spin-3 state overlap under the peak. They are
$D^*_{s1}(2860)$ with mass and width as
$M(D^*_{s1}(2860))=(2859\pm 12\pm 6\pm 23){\rm MeV},
\,\Gamma(D^*_{s1}(2860))=(159\pm 23\pm 27\pm 72){\rm MeV}$
\cite{Aaij:2014xza} and
$D^*_{s3}(2860)$ with mass and width as
$M(D^*_{s3}(2860))=(2860.5\pm 2.6\pm 2.5\pm 6.0){\rm MeV},
\,\Gamma(D^*_{s3}(2860))=(53\pm 7\pm 4\pm 6){\rm MeV}
$\cite{Aaij:2014baa}. Based on the new data  Godfrey and Moats
suggest that \cite{Godfrey:2013aaa} $D^*_{s1}(2860)$ and
$D^*_{s3}(2860)$ should be identified as $1^3D_1(c\bar s)$ and
$1^3D_3(c\bar s)$. Previously
$D^*_{s1}(2710)$\cite{Brodzicka:2007aa} was measured and its mass
and width are $M(D^*_{s1}(2710))=(2709\pm 4){\rm MeV},
\,\Gamma(D^*_{sJ}(2710))=(117\pm13){\rm MeV}$. It
was assigned to be $1^3D_1$ or $2^3S_1$ or their
mixture \cite{Godfrey:2013aaa,Close:2006gr,Zhang:2006yj}. Obviously
the $1^3D_1$ assignment of $D^*_{s1}(2710)$ conflicts with
the LHCb's new observation, because the $1^3D_1$ state of $c\bar s$
is occupied by $D^*_{s1}(2860)$, so there is
no room to accommodate  $D^*_{s1}(2710)$. Therefore one can
conjecture that as long as $D^*_{s1}(2860)$ is in the $1^3D_1$
state, $D^*_{s1}(2710)$ should be regarded as a $2^3S_1$
state or others\cite{Godfrey:2013aaa,Godfrey:2014fga}. Since all resonances
$D^*_{s1}(2860)$ and $D^*_{s3}(2860)$ and $D^*_{s1}(2710)$ have been undoubtedly
reconstructed in the hadronic processes under investigation,
the best channels to determine their quantum identities are their
respective strong
decays\cite{Godfrey:2013aaa,Wang:2014jua,Song:2014mha} which are in fact
the dominant ones. However, on other aspect, one still has a
chance to observe the resonances in their electromagnetic decays where
excited states transit into ground states by emitting a photon.
Especially the calculation on the electromagnetic decays is more
reliable.  In Ref.\cite{Li:2009qu} the authors study the radiative
decays of $D_s^*$$(1^3D_1)$ and $D_s^*$$(3^3D_1)$ into a P-wave
$c\bar s$ meson. In this paper we will study the radiative decay
of a D-wave meson into an S-wave $c\bar s$ meson. The results may
help us to determine the quantum number of these particles in
addition to the studies via strong processes.

In this work, we employ the light-front quark model(LFQM)
to estimate the branching ratios. This relativistic model
has been thoroughly discussed in literatures
\cite{Jaus:1999zv,Cheng:2003sm} and applied to study hadronic
transition processes\cite{Wei:2009nc,Ke:2012wa,Ke:2013yka}. The
results obtained in this framework qualitatively agree with the
data for all the concerned processes.

In conventional LFQM  the radiative decay of a  $1^{--}$ (S-wave)
meson into a $0^{-+}$ meson  was evaluated  \cite{Choi:2007se} and
the same formula can also be generalized to the covariant LFQM
\cite{Hwang:2006cua}. In our earlier
papers\cite{Ke:2010vn,Ke:2011jf,Ke:2013zs} we studied radiative
decays of some mesons in covariant LFQM and now we will
concentrate our attention to the radiative decays of   $1^{--}$
(D-wave) mesons to $0^{-+}$ mesons. The results would be useful
for confirming the identities of the aforementioned mesons. Since
the Lorentz structure of the vertex functions of D-wave is the
same as that of S-wave \cite{Ke:2011mu}, the formulas for decays
of the $1^{--}$ D-wave mesons can be simply obtained by replacing
several functions which were used for the decays of the $1^{--}$
S-wave mesons.

This paper is organized as following: after this introduction, we derive the theoretical formulas in next section
where we also present relevant formulas given in literatures,
and then in Sec. III, we present our numerical results along with all inputs which are needed for the numerical computations. In the
last section we draw our conclusion and make a brief discussion.

\section{The formulas for the radiative decay of $1^{--}$
meson in LFQM}
\begin{center}
\begin{figure}[htb]
\begin{tabular}{cc}
\scalebox{0.5}{\includegraphics{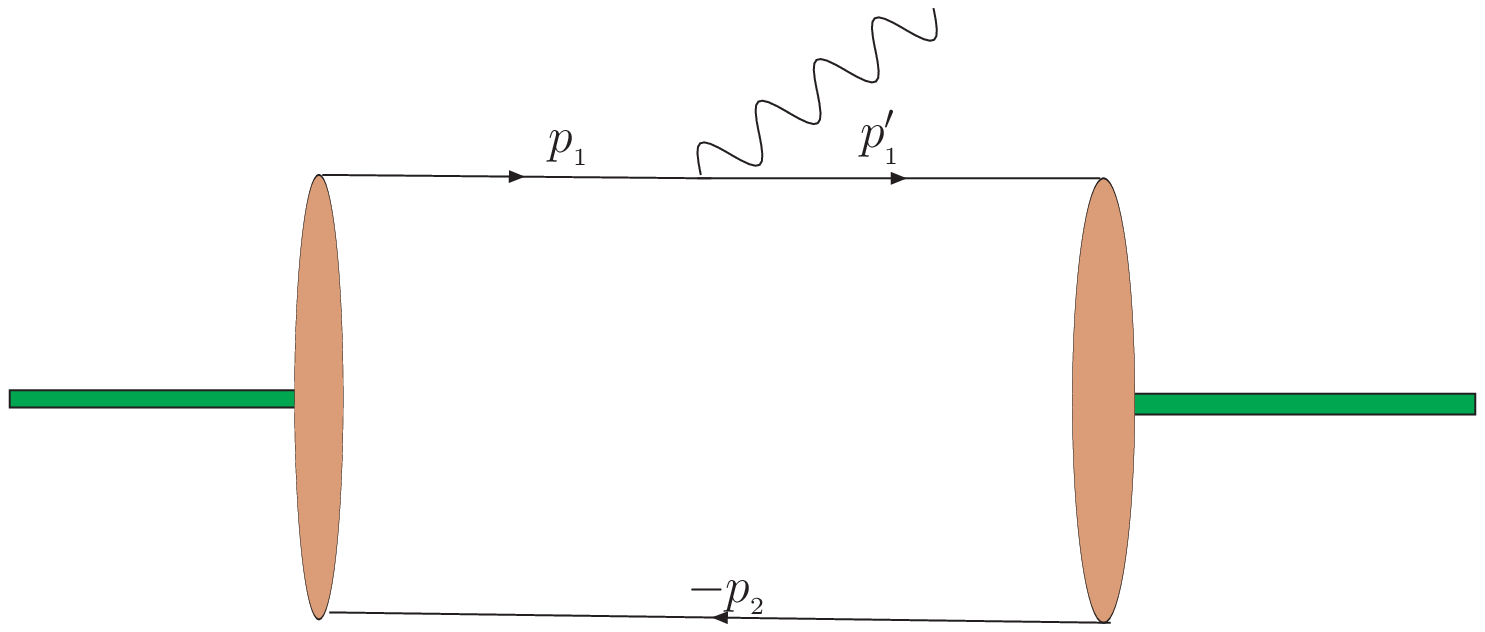}}&\raisebox{-2.em}
{\scalebox{0.5}{\includegraphics{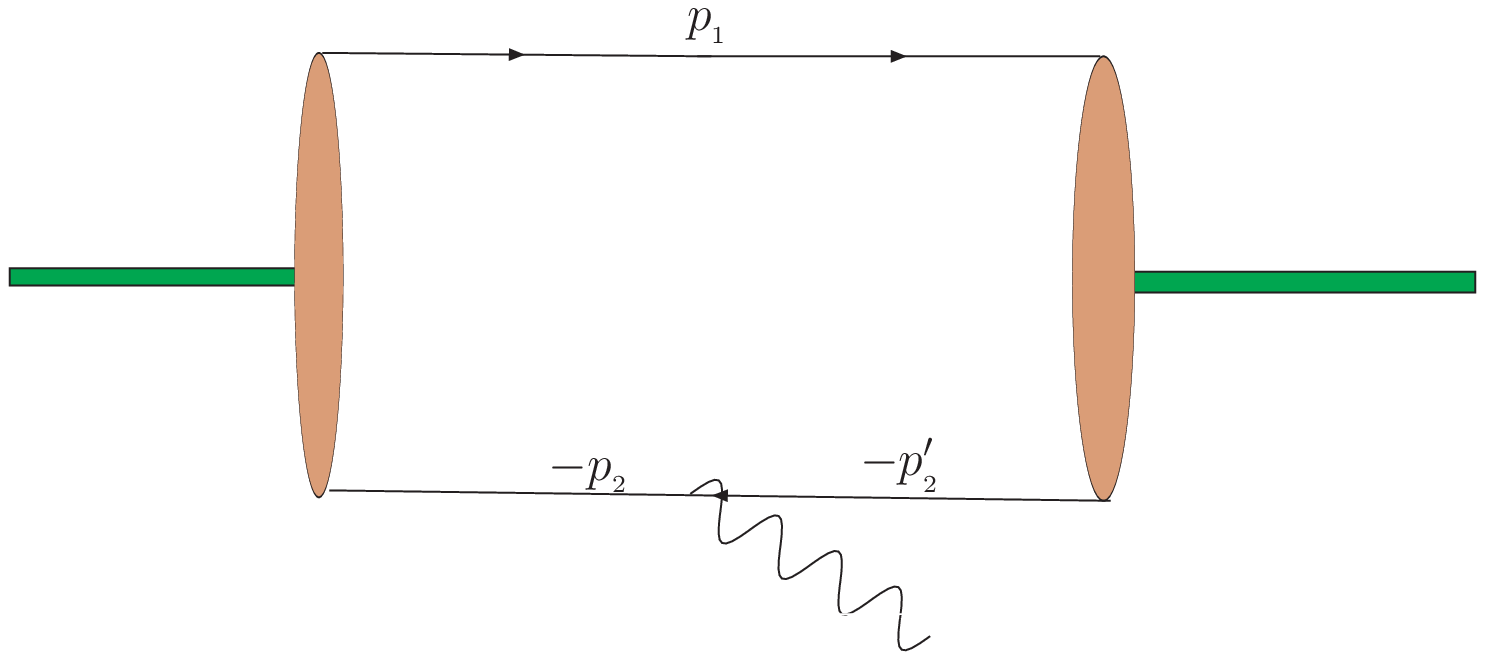}}}
\end{tabular}
\caption{Feynman diagrams depicting the radiative decay
.\label{fig:LFQM}}
\end{figure}
\end{center}

In the light front quark model, the transition matrix elements for
the decay of $1^{--}(V)\rightarrow 0^{-+}(P) \gamma$ were examined
(Fig.\ref{fig:LFQM}) and   the form factor $\mathcal{ F}_{V\to
P}(q^2)$ can be expressed as \cite{Choi:2007se}:
\begin{eqnarray}\label{21}
\mathcal{ F}_{V\to P}(q^2)= e_1I(m_1,m_2,q^2) + e_{2}
I(m_2,m_1,q^2),
\end{eqnarray}
where $e_{1}$ and $e_{2}$  are the electrical charges of charm
and strange quarks, $m_1=m_c$, $m_2=m_s$ and
\begin{eqnarray}\label{22}
I(m_1,m_2,q^2)&=&\int^1_0 \frac{dx}{8\pi^3}\int d^2{\bf p}_\perp
\frac{\phi\phi'\biggl\{{\cal A} + \frac{2} { w_{_{V}}} [{\bf
p}^2_\perp - \frac{({\bf p}_\perp\cdot{\bf q}_\perp)^2}{{\bf
q}^2_\perp}]\biggr\}}{x_1\tilde{M_0}\tilde{M'_0}}
\nonumber\\&=&N_c\int^1_0 \frac{dx}{4\pi^3}\int d^2{\bf p}_\perp
\frac{h_{^3{S_1}}h'_{P}\biggl\{{\cal A} + \frac{2} {
w_{_{^3{S_1}}}} [{\bf p}^2_\perp - \frac{({\bf p}_\perp\cdot{\bf
q}_\perp)^2}{{\bf
q}^2_\perp}]\biggr\}}{x_1^2x_2(M^2-M_0^2)(M'^2-{M'_0}^2)} ,
\end{eqnarray}
where
$h_{{^3{S_1}}}=h_P=(M^2-M_0^2)\sqrt{\frac{x_1x_2}{N_c}}\frac{1}{\sqrt{2}\tilde{M}_0}\phi,$
$w_{_{^3{S_1}}}=M_0+m_1+m_2,$ ${\cal A}=x_2m_1+x_1m_2$ and
$x=x_1$.  It is noted that the $1^{--}$ meson in
Ref.\cite{Choi:2007se,Hwang:2006cua} just refers to $^3S_1$ state.
The other variables in Eq. (\ref{22}) are presented in the Appendix.

Obviously, a $1^{--}$ meson may be in a $^3D_1$ state
or a $^3S_1$ state or their mixture.

In Ref.\cite{Ke:2011mu} the vertex function for $^3D_1$ states was deduced and
its Lorentz structure is the same as that of $^3S_1$ state, so
Eq.(\ref{22}) is also valid for the radiative decay of $^3D_1$ through replacing the functions
$h_{^3S_1}$ and $w_{^3S_1}$ by
$$h_{(^{3}D_1)}=-(M^2-M_0^2)\sqrt{\frac{x_1x_2}{N_c}}\frac{1}{\sqrt{2}\tilde{M_0}}
\frac{\sqrt{6}}{12\sqrt{5}M_0^2\beta^2}[M_0^2-(m_1-m_2)^2][M_0^2-(m_1+m_2)^2]\phi,$$
$$w_{(^{3}D_1)}=\frac{(m_1+m_2)^2-M_0^2}{2M_0+m_1+m_2}.$$

The decay width is\cite{Choi:2007se}
\begin{eqnarray}\label{23}
\Gamma(V\rightarrow
P+\gamma)=\frac{\alpha}{3}\bigg[\frac{m_{V}^2-m_{P}^2}{2m_{V}}\bigg]^3
\mathcal{ F}^2_{V \to P}(0).
\end{eqnarray}

\section{Numerical results}
Before we carry out our numerical computations for evaluating the
branching ratios of the D-wave mesons, we need to determine a
nonperturbative  parameter $\beta$ which exists in the wave
function, in a proper way. In Ref.\cite{Cheng:2003sm} the authors
suggested that via calculating the decay constant of the ground
state one can determine $\beta$. Alternatively, we also can get
the value of  $\beta$ by  fitting the spectra of the relevant
mesons as done in Ref.\cite{Choi:2007se}. In this work we follow
the first scheme. With the averaged decay branching ratio of
$D_s\rightarrow\mu\nu_\mu$ $(5.56\pm0.25)\times
10^{-3}$\cite{PDG12} one obtains its decay constant as
$f_{D_s}=(247\pm6)$ MeV. Then using the Eq.(6) in
Ref.\cite{Hwang:2006cua} $\beta$ is fixed as $(0.534\pm0.015)$
GeV$^{-1}$ when we set $m_c=1.4$GeV,
$m_s=0.37$GeV\cite{Cheng:2003sm} and
$m_{D_s}=1.9685$GeV.\cite{PDG12}
\subsection{The radiative decays
of $D^*_{s1}(2860)$ and $D^*_{s3}(2860)$}

In our numerical computations we adopt the assumption that  $D^*_{s1}(2860)$
and $D^*_{s3}(2860)$ are $1^3D_1(c\bar s)$ and $1^3D_3(c\bar s)$
respectively.

\begin{center}
\begin{figure}[htb]
\begin{tabular}{cc}
\scalebox{0.8}{\includegraphics{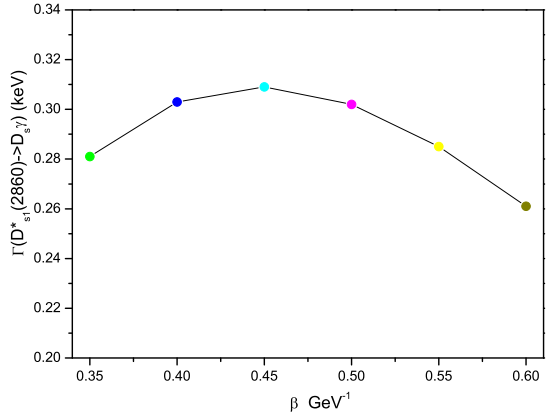}}
\end{tabular}
\caption{$\Gamma(D^*_{s1}(2860)\to D_s\gamma)$ dependence on
$\beta$ .\label{dp}}
\end{figure}
\end{center}

Using the parameters we calculate the form factor  $\mathcal{
F}(0)$  for $D^*_{s1}(2860)\to D_s\gamma$ which is
$(0.0168\pm0.0002)$ GeV$^{-1}$. The decay width
$\Gamma(D^*_{s1}(2860)\to D_s\gamma)$ is ($0.291\pm0.006$) keV.
Comparing with the total width the value is rather small, namely the branching ratio is small, but one still
has a chance to measure it in more accurate experiments. To explore its dependence
on the parameter $\beta$ we vary $\beta$ from 0.35 GeV$^{-1}$ to
0.6 GeV$^{-1}$. The results are depicted in Fig.\ref{dp}. One can
notice that the result is not sensitive to the value of $\beta$ after all.

Since the vertex function of the $^3D_3$ state is more complicated we are
not going to directly deduce the transition matrix elements for the radiative decays in this framework.
Instead, we would take an approximate but reasonable scheme to estimate the radiative decay width of  $^3D_3$.
Namely, one obtains the rate of $^3D_3$ radiative decay in
terms of that of the $^3D_1$ radiative decay. Under the nonrelativistic
approximation the authors of Ref.\cite{Novikov:1977dq} presented  a formula to
calculate the widths for the $M1$ transition as
\begin{equation}\label{Gamma}
 \Gamma(i\rightarrow f\gamma)=\frac{\alpha}{3}(\frac{e_c}{m_c}-\frac{e_{\bar s}}{m_s})^2{E_\gamma}^3(2J_f+1)|\langle f|j_0(kr/2)|i\rangle|^2.
\end{equation}\
If we ignore the  spin-orbit coupling term in the potential which
results in the fine-structure of spectra, the wave functions of
$D^*_{s1}(2860)$ and $D^*_{s3}(2860)$ obtained by solving the
Sch\"{o}rdinger equation would be the same because they have the
same orbital angular momentum and intrinsic spin, thus we would
naturally get $\langle D_s|j_0(kr/2)|D_s(^3D_3)\rangle=\langle
D_s|j_0(kr/2)|D_s(^3D_1)\rangle$. Since the mass of
$D^*_{s1}(2860)$ is close to that of  $D^*_{s1}(2860)$, it hints
that the contributions of the spin-orbit coupling term to spectra
and wave function are less important. By including all factors, it
is straightforward to estimate $\Gamma(D_s(^3D_3)\rightarrow
D_s\gamma)\approx\Gamma(D_s(^3D_1)\rightarrow D_s\gamma)$.

\subsection{The radiative decay of $D^*_{s1}(2710)$ }
\begin{table}
\caption{The form factor for $D^*_{s1}(2710)\to D_s$ .}
\label{tab:expect}
\begin{tabular}{c|c|c|c}\hline
 ~~~~~~~~   &  D-wave  &
 S-wave(1)& S-wave(2)\\\hline
 $\mathcal{ F}(0)$ (GeV$^{-1}$)   &$-0.0168\pm0.0002$    &  $0.099\pm0.001$      & $0.112\pm0.001$   \\
 $\Gamma$ (keV)   &$0.179\pm0.004$    & $6.18\pm0.07$     & $8.00\pm0.02$    \\
\hline
\end{tabular}
\end{table}
After $D^*_{s1}(2710)$ was found, a lot of work has been done to
investigate its identity.  In Ref.\cite{Godfrey:2014fga} the authors
suggested that $D^*_{s1}(2710)$ should be a $2^3S_1$ state, rather than a
$1^3D_1$. To be more open, here let us assume  $D^*_{s1}(2710)$ to be respectively a $2^3S_3$ state or a $1^3D_1$
state and under the different assumptions, we calculate its radiative decay width. The results are listed
in table \ref{tab:expect}. For the S-wave state ($2^3S_3$) we employ the conventional
wave function (S-wave(1)) and modified wave function (S-wave(2))
which was discussed in Ref.\cite{Ke:2010vn}. Then we continue to calculate the rate of radiative decay
of the D-wave state in the aforementioned approximation.

One would notice that  there exists
a huge gap between the S-wave and D-wave cases.

\begin{center}
\begin{figure}[htb]
\begin{tabular}{cc}
\scalebox{0.8}{\includegraphics{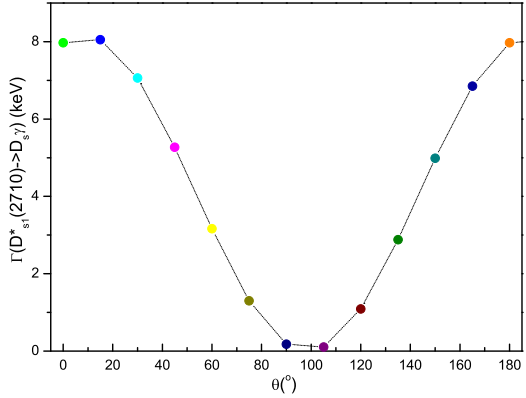}}
\end{tabular}
\caption{dependence of $\Gamma(D^*_{s1}(2710)\to D_s\gamma)$  on the
mixing angle $\theta$.\label{mix}}
\end{figure}
\end{center}

If we assume that $D^*_{s1}(2710)$ is the mixture of $2^3S_3$
and $1^3D_1$ i.e. $|D_{(s1)}(2710)\rangle={\rm
Cos}\theta|2^3S_3\rangle-{\rm
Sin}\theta|1^3D_1\rangle$\cite{Li:2009qu}, using the values of
$\mathcal{F}(0)$ given in table (\ref{tab:expect}) the
corresponding radiative decay width is re-calculated. In Fig.\ref{mix} the dependence of the decay width on
the mixing angle $\theta$ is depicted where the modified
wavefunction is used for the 2S state.

In Ref.\cite{Li:2009qu} the authors
studied  $\Gamma(D_s(2710)\rightarrow D_{s2}(2573)\gamma)$,
$\Gamma(D_s(2710)\rightarrow D_{s0}(2317)\gamma)$,
$\Gamma(D_s(2710)\rightarrow D_{s1}(2460)\gamma)$
 and $\Gamma(D_s(2710)\rightarrow D_{s1}(2536)\gamma)$ which are $0.09\sim
 0.12$keV, $7.80\sim 7.97$ keV, $1.47\sim 1.56$ keV and $0.27\sim 0.29$ keV respectively.
The above cited estimates are about radiative decays of $D_s(2710)$ into a P-wave meson plus
a photon, as we noted that for finally identifying the quantum numbers of $D_s(2710)$, the
decay mode under investigation: $D_s(2710)\rightarrow
D_{s}(1963)\gamma$ which is $D_s(2710)$ decaying into a S-wave meson plus a photon, is not less important.

\section{Summary}
In this work we study the radiative decay of $D^*_{s1}(2860)$,
$D^*_{s3}(2860)$ and $D^*_{s1}(2710)$ respectively in terms of
LFQM. Assuming $D^*_{s1}(2860)$, $D^*_{s3}(2860)$ to be $1^3D_1$
and $1^3D_3$ states, we obtain their partial widths. Our estimates
on $\Gamma(D^*_{s1}(2860)\to D_s\gamma)$ and
$\Gamma(D^*_{s1}(2860)\to D_s\gamma)$  are approximately $0.291$
keV. The estimated branching ratios of the radiative decays of
$D^*_{s1}(2860)$ and $D^*_{s3}(2860)$  are about 1.9$\times
10^{-6}$ and 5.8 $\times 10^{-6}$. By the achieved integrated
luminosity at LHCb (3.0 fb$^-1$), the LHCb
Collaboration\cite{Aaij:2014baa} collected 12450 $B_s^0\rightarrow
\bar{D^0}K^-\pi^+$ samples where only a part of the events concern
$D^*_{s1}(2860)$ and $D^*_{s3}(2860)$. Their radiative decays have
not been observed yet due to the small database for $D^*_s(2860)$.
Indeed we need longer time and higher luminosity to observe the
radiative decays $\Gamma(D^*_{s1}(2860)\to D_s\gamma)$ and
$\Gamma(D^*_{s1}(2860)\to D_s\gamma)$.

Though the fractions of the radiative decays are small, they
have clear signal to be observed from the background, therefore the advantage of
detecting those modes is obvious. Thus we expect our experimental
colleagues to carry out accurate experiments to measure them.

Concerning $D^*_{s1}(2710)$, as discussed
in the introduction, if $D^*_{s1}(2860)$ and $D^*_{s3}(2860)$ are confirmed to be
the D-wave $D_s$ meson,  $D^*_{s1}(2710)$ cannot be a pure 1D-wave $c\bar s$ system, we calculate
its radiative decay rate by assuming two possible assignments: $2^3S_3$ or $1^3D_1$ respectively. Our numerical results show that if
it is a $2^3S_3$ state the corresponding branching ratio is about 5.2
$\times 10^{-5}\sim$6.7 $\times 10^{-5}$, instead while it is
$1^3D_1$, the corresponding rate is around 1.5 $\times
10^{-6}$. There is an obvious gap between the estimated rates for the two assignments.

Because the LFQM is a relativistic model
and its validity is widely recognized due to its success for explaining the available data for hadronic decays of heavy mesons,
we may believe that the numerical results obtained in this framework is trustworthy, at most they could only decline from the real values
by a small factor less than 2 which was confirmed by other phenomenological studies in terms of the same model.
The possible uncertainties are incurred by the inputs. Even so, the results could help identifying the quantum
numbers since in the two cases the resultant ratios of $\Gamma(D^*_{s1}(2710)\to D_s\gamma)$
are apparently apart.

No doubt, the final decision will be made by the future precise measurements. Our work only indicates the importance of
studying the radiative decays because of their obvious advantage and strongly suggest to search such decay modes at the coming
super-BELLE or next run of LHCb, even the expected ILC.

\section*{Acknowledgement}

This work is supported by the National Natural Science Foundation
of China (NNSFC) under the contract No. 11375128.

\appendix

\appendix

\section{Notations}

Here we list some variables appearing in the context.  The
incoming  meson in Fig. \ref{fig:LFQM} has the momentum
$P=p_1+p_2$ where $p_1$ and $p_2$ are the momenta of the off-shell
quark and antiquark and
\begin{eqnarray}\label{app1}
&& p_1^+=x_1P^+, \qquad ~~~~~~p_2^+=x_2P^+, \nonumber\\
&& p_{1\perp}=x_1P_{\perp}+p_\perp, \qquad
 p_{2\perp}=x_2P_{\perp}-p_\perp,
 \end{eqnarray}
with  $x_i$ and $p_\perp$ are internal variables and $x_1+x_2=1$.

The variables $M_0$ and $\tilde {M_0}$ are defined as
\begin{eqnarray}\label{app2}
&&M_0^2=\frac{p^2_\perp+m^2_1}{x_1}+\frac{p^2_\perp+m^2_2}{x_2},\nonumber\\&&
\tilde {M_0}=\sqrt{M_0^2-(m_1-m_2)^2},\nonumber\\&&
\phi(1S)=4(\frac{\pi}{\beta^2})^{3/4}\sqrt{\frac{dp_z}{dx_2}}{\rm
 exp}(-\frac{p^2_z+p^2_\perp}{2\beta^2}),\nonumber\\&&
\phi(2S)=4\Big(\frac{\pi}{\beta^2}\Big)^{3/4}\sqrt{\frac{\partial
p_z}{\partial
x}}{\exp}\Big(-\frac{1}{2}\frac{p^2_z+p^2_\perp}{\beta^2}\Big)
\Big(3
-2\frac{p^2_z+p^2_\perp}{\beta^2}\Big)\frac{1}{\sqrt{6}}\nonumber\\&&
\phi_M(2S)=4\Big(\frac{\pi}{\beta^2}\Big)^{3/4}\sqrt{\frac{\partial
p_z}{\partial
x_2}}{\exp}\Big(-\frac{{2}^\delta}{2}\frac{p^2_z+p^2_\perp}{\beta^2}\Big)
\Big(a_2 -b_2\frac{p^2_z+p^2_\perp}{\beta^2}\Big).
 \end{eqnarray}
with $p_z=\frac{x_2M_0}{2}-\frac{m_2^2+p^2_\perp}{2x_2M_0}$,
$\delta=1/1.82$, $a_2=1.88684$ and $b_2=1.54943$.


\end{document}